\documentclass{myelsart}
\usepackage{graphicx}
\usepackage{natbib}
\usepackage{amsbsy}

\newcommand{\be}{\begin{equation}}

\begin{document}

\runauthor{U. Becciani et al.}
\begin{frontmatter}
\title{VisIVO - Integrated Tools and Services for Large-Scale Astrophysical Visualization}

\author[OACT]{U. Becciani,}
\author[OACT]{A. Costa,}
\author[OACT]{V. Antonuccio-Delogu,}
\author[OACT]{G. Caniglia,}
\author[UNICT]{M. Comparato,} 
\author[CINECA]{C. Gheller,}
\author[PORT]{Z. Jin,}
\author[PORT]{M. Krokos,}
\author[OACT]{P. Massimino}

\address[OACT]{INAF - Osservatorio Astrofisico di Catania, Italy}
\address[CINECA]{CINECA - Casalecchio di Reno, Italy}
\address[UNICT]{Dipartimento di Fisica e Astronomia, University of Catania, Italy}
\address[PORT]{School of Creative Technologies, University of Portsmouth, United Kingdom}

\begin{abstract}
VisIVO  is an integrated suite of tools and services specifically designed for the Virtual Observatory. This suite constitutes a software framework for effective visual discovery in currently available (and next-generation) very 
large-scale astrophysical datasets. VisIVO consists of VisiVO Desktop - 
a stand alone application for interactive visualization on standard PCs, VisIVO Server - a grid-enabled platform for high performance visualization and VisIVO Web - a custom designed web portal supporting services based on the VisIVO Server functionality. 
The main characteristic of VisIVO is support for high-performance, 
multidimensional visualization of very large-scale astrophysical datasets. 
Users can obtain meaningful visualizations rapidly while preserving full and intuitive control of the relevant visualization parameters. This paper focuses on  newly developed integrated tools in VisIVO Server allowing intuitive visual discovery with 3D views being created from data tables. VisIVO Server can be installed easily on any web server with a database repository. We discuss briefly aspects of our implementation of VisiVO Server on a computational grid and also outline the functionality of the services offered by VisIVO Web. Finally we conclude with a summary of our work and pointers to future developments.
\end{abstract}
\end{frontmatter}

\section{Introduction}
An essential part of astrophysical research is the necessity to employ computer graphics and scientific visualization tools for displaying appropriately multi-dimensional data plots and images either from real-world observations or from modern, highly-complex numerical simulations. 
The latest generation of graphics and visualization software tools offer the astrophysical community robust instruments for data analysis and advanced data exploration by using:
\begin{enumerate}
\item {\it High Performance Computing} and {\it Multithreading} to exploit multi-core CPUs and emerging powerful graphics boards, thus allowing real-time interaction with large-scale datasets.

\item {\it Interoperability} so as to allow different applications, each specialized for different purposes, to operate on shared datasets.

\item {\it Collaborative Workflows} so that several users can work simultaneously - perhaps at different geographical locations - on identical datasets exchanging information and visualization experiences.

\end{enumerate}

Astrophysicists are currently witnessing an unprecedented growth in the quality and quantity of datasets coming from real-world observations and numerical simulations. For example, the increasing availability of high performance and grid computing facilities \citep{BeccianiGrid} \citep{ube} has given the possibility to perform numerical simulations of several dimensions. Also current, such as the Sloan Digital Sky Survey (SDSS) \citep{SDSS} \citep{WWWSDSS}, and next-generation sky surveys, such as the Low Frequency Array (LOFAR) \citep{LOFAR}, the Square Kilometre Array (SKA) \citep{SKA}, the Large Synoptic Survey Telescope System (LSST)\citep{LSST}, and the Dark Energy Survey (DES) \citep{DES}, are planned to collect large amounts of raw data eventually resulting in several hundreds of terabytes of data in tabular form. These developments place the emphasis on the problems associated with data post-processing. For storage, very large-scale distributed databases will be required and scientists will need to interact with them effectively using very high performance computer graphics and scientific visualisation tools as simply downloading will be prohibitively costly. The emerging need is not only to provide data analysis and exploration tools that can run on standard PCs satisfactorily, but also to offer appropriate interoperable tools that can run on servers and standard PCs simultaneously.\\
This paper describes VisIVO  Server \citep{visivoserver}  a new platform for astrophysical visualization of large-scale datasets that can be easily installed on any computing server and can handle datasets in cooperation with VisIVO Desktop \citep{wwwvisivodk}, our previously developed, stand-alone visualization application for standard PCs. Section 2 includes a short discussion on existing visualization tools for astrophysics
 and outlines the VisIVO Desktop functionality. The VisIVO Server is discussed in detail in section 3, including a presentation of its main components namely, VisIVO Importer, VisIVO Filters and VisIVO Viewer. Section 4 discusses aspects of our current grid installation of VisIVO Server while VisIVO Web, a portal for providing VisIVO Server services, is outlined in section 5. Finally section 7 presents a summary of our work and includes pointers to future developments.

\section{Background}
Traditionally the common practice among astrophysicists is to employ a plethora of individually created, autonomous applications for data analysis and exploration. However this scenario is not applicable to modern large-scale datasets and over the last few years a generation of software frameworks has emerged, e.g. Aladin \citep{Aladin}, that is  an interactive software sky atlas allowing the user to visualize digitized astronomical images, superimpose entries from astronomical catalogues or databases, and interactively access related data and information from archives,  and TopCat \citep{Topcat}  an interactive graphical viewer and editor for tabular data. Other examples of this new generation of software are 3D slicer \citep{wwwSlicer} that is capable of displaying volumes and user-defined cut-throughs, SPLASH \citep{Splash}  for visualisation of numerical simulations and VisIt  \citep{wwwVisit} which is a free parallel visualisation and graphical analysis tool for viewing scientific datasets and contains a rich set of visualisation features e.g. display of scalar and vector fields defined on two or three dimensional structured and unstructured meshes.\\
We have previously developed VisIVO Desktop, a stand-alone visualization application specifically designed for interactive exploration of large-scale multidimensional astrophysical datasets. It was the first component within the VisIVO family of visualization tools and further information can be found in \cite{visivo}. Many features of VisIVO Server are derived from VisIVO Desktop. The VisIVO Desktop application is wrapped around the  Multimod Application Framework \citep{maf}, which is an open-source software framework for rapid prototyping of data analysis and visualization applications. 
\begin{figure}
\begin{center}
\includegraphics[width=12.0cm]{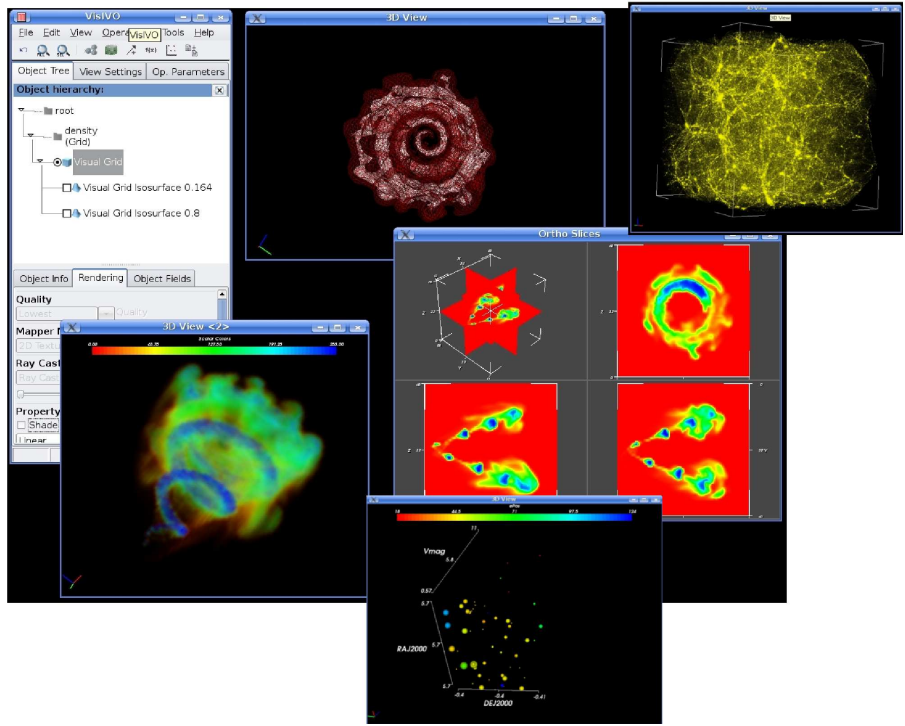}
\caption{A typical user session with VisIVO Desktop illustrating (from left to right and from top to bottom): a) the user interface window, b) an isosurface rendering, c) a rendering of unstructured data points from a cosmological simulation, d) a volume rendering and cutting planes visualisation of structured data from a fluidodynamical simulation and finally e) a geometrical  representation of data points from astrophysical catalogues.}
\label{des}
\end{center}
\end{figure}
The current release (version 1.5) supports several popular astronomical data formats (e.g. VO and FITS tables) and hardware platforms (e.g. Windows, Mac OSX and many Linux distributions). VisIVO  application handles observational and simulated datasets, especially focusing on multiple dimensions, such as catalogues or computational meshes. The data is represented either as points or volumes and can be visualized using advanced rendering - such as isosurfacing and volume rendering algorithms - or even creation and visualization of vectors by combining a user-defined selection of  scalar values (Fig. \ref{des}).\\ 
The next section discusses in detail the VisIVO Server functionality, that constitutes the latest development in the VisIVO project for exploration of large-scale datasets.
In the following, we report on performance tests of some VisIVO Server modules as an indication for users interested in running VisIVO Filters on a local or remote server. All our performance tests were executed on  an  AMD Opteron 2218 rev. F dual-core processor with a clock rate of 2,6 GHz and 2 GB of DDR2 RAM.\\

\section{VisIVO Server}
VisIVO Server is a suite of software tools for creating customized views of 3D renderings from astrophysical data tables.  These tools are founded on the VisIVO Desktop functionality  and support the most popular Linux based platforms (e.g. www.ubuntu.com). Their defining characteristic is that no fixed limits are prescribed regarding the dimensionality of data tables input for processing, thus supporting very large-scale datasets. VisIVO Server consists of three core components: {\bf VisIVO Importer}, {\bf VisiVO Filter} and {\bf VisIVO Viewer} respectively.\\
To create customized views of 3D renderings from astrophysical data tables, a two-stage 
process is employed. First, VisIVO Importer is utilized to convert user datasets into VisIVO Binary Tables (referred to as VBTs in the remainder of this article). Then, VisIVO Viewer is invoked to display customized views of 3D renderings. As an example, consider displaying views from only three columns of an astrophysical data table supplied in ascii form, say col\_1, col\_2 and col\_3; this is achieved by executing the following VisIVO Server commands:
\begin{verbatim}
VisIVOImporter --fformat ascii --out VBT.bin UserDataSet.txt
VisIVOViewer --x col_1 --y col_2 --z col_3  VBT.bin
\end{verbatim}
where UserDataSet.txt is the user input ascii data table and VBT.bin is the binary VBT corresponding to it, created by the VisIVOImporter command.\\
VisIVO Filter is a suite of filters used to generate new VBTs from existing VBTs, e.g. consider constructing a new version of a user dataset through randomized sub-sampling in order to fit it in the available RAM. The remainder of this section describes the functionality of the VisIVO Server core components.\\
A VBT is a highly-efficient data representation used by VisIVO Server internally. A VBT is realized through a header file (extension .bin.head) containing all necessary metadata, and a raw data file (extension .bin) storing actual data values. The header contains information regarding the overall number of fields and number of points for each field (for point datasets) or the number of cells and relevant mesh sizes (for volume datasets). The raw data file is typically a sequence of values, e.g. all X followed by all Y values. The header file contains the following fields:
\begin{verbatim}
float (double)
n1
n2 [ CellX CellY CellZ DX DY DZ ]
little | big 
X
Y
Z
Vx
Vy
Vz
\end{verbatim}
where
\begin{itemize}
\item {\bf float} (double) is the data type of the storage variables used; 
\item {\bf n1} denotes the number of columns  in the VBT;
\item {\bf n2} denotes the number of rows in the VBT;
\item {\bf CellX, CellY, CellZ, DX, DY, DZ} are given only if the VBT represents volumetric datasets. In that case $CellX$, $CellY$ and $CellZ$ represent the mesh geometry, while $DX$, $DY$ and $DZ$ represent the x, y and z size of volumetric cells.
\item {\bf little | big} denotes  the endianism employed in the VBT. 
\end{itemize}

\subsection{VisIVO Importer}
VisIVO Importer converts user-supplied datasets into VBTs without imposing any limits on sizes or dimensionality. VBTs are typically employed by VisIVO Filter modules for data processing and by VisIVO Viewer for the final display. The general syntax needs to specify the input data format and the user filename.
The filename can be a local or remote file   to be  converted into the VBT. If a remote filename is given (starting with http://, sftp:// or ftp://) the remote file is downloaded automatically using the curl library \citep{Curl}. However if a username and password are specified, the prescribed username and password are  employed for remote access.\\ 
The current version of VisIVO Importer supports conversion from several popular formats as follows:
\begin{itemize}
\item {\bf ASCII} and {\bf CSV} - ASCII files normally contain values for a number of variables organised in columns. The columns are typically separated by white space characters, e.g. spaces or tabs. VisIVO Importer expects the first row of ASCII files to contain the names of the corresponding variables. 
CSV is a delimited data format that has fields/columns separated by the comma character and records/rows separated by newlines. The CSV file format does not require a specific character encoding, byte order, or line terminator format.
\item {\bf VOTables} - The VOTable format is an XML standard for the interchange of data represented as a set of tables \citep{VOTable}. A table is an unordered set of rows, each having a uniform format, as specified in the table metadata information. Each row in a table is a sequence of table cells, and each of these contains either a primitive data type or an array of such primitives.
\item {\bf FLY} - FLY is code that uses the tree N-body method, for three-dimensional self-gravitating collisionless systems evolution \citep{2007CoPhC.176..211B}. FLY is a fully parallel code based on the tree Barnes-Hut algorithm; periodical boundary conditions are implemented by means of the Ewald summation method. 
FLY is founded on the one-side communication paradigm for sharing data among processors that access remote private data while avoiding any kind of synchronism. 
\item {\bf FITS Tables} - FITS is a codification into a formal standard, by the NASA/Science Office of Standards and Technology, of the FITS rules endorsed by the IAU. FITS supports tabular data with named columns and multidimensional rows \citep{Fits}. 
\item {\bf Gadget} - It is a freely-available code for cosmological N-body/SPH simulations on massively parallel computers with distributed memory \citep{Gadget}. GADGET uses an explicit communication model that is implemented with the standardized MPI communication paradigm. 
\item {\bf Raw Binary} and {\bf Raw Grid} - Raw files are simply  binary memory dumps of the  for data points.  The content of the Raw Binary data points file is a sequence of x,y and z coordinates for each point, then a sequence of fields, one scalar for each data point. Raw Grid contains only one quantity: the content of a volume file is a sequence of values, one value for each mesh point. 
\item {\bf TVO XML} - The TVO XML format is essentially the VOTable format enhanced with all necessary information to allow data downloading. It is used to describe the content of a generic binary file, and it contains the reference to the input files.  If  a filename starts with http://, sftp:// or ftp:// the remote file is  downloaded using the CURL library. The TVO XML document is currently under revision of the TVO Interest group in the IVOA framework \citep{costa}.  
\end{itemize}
The operation of VisIVO Importer is highly optimised requiring in most cases a short period of time (a number of seconds) even for large-scale datasets. For example, no more than 30 secs are necessary for importing 30 million binary Gadget elements.

\subsection{VisIVO Filters}
VisIVO Filter is a  collection of data processing modules to modify a VBT or to create a new VBT from existing VBTs. The filters support a  range of operations such as scalar distribution,  mathematical operations, selections of regions, decimation, randomization and so on. The selection and randomization operations are of particular importance as they are typically employed for constructing reduced VBTs  so that they can be used directly by VisIVO Viewer.
The general syntax is 
\begin{verbatim}
VisIVOFilters --op filterOpCode <options> [--file] InputFile
\end{verbatim}
where  $filterOpCode$ selects a specific operation to be processed (e.g. randomization), $options$ is a set of flags
 related to the specific selected operation and $InputFile$ is an input VBT file or a list of VBT files.
The following subsections describe in detail the operation of the main VisIVO Filter modules.

\subsubsection{Randomization}  
This operation creates a new VBT table as a random subsample from the input table. The operation allows the user to produce a statistically identical version of the original table in order to allow  efficient data analysis, or to visualize data using VisIVO Viewer. A similar VisIVO Filter operation ({\it decimator}, see section 3.2.10) allows to select a fixed subsets of input datasets.

\subsubsection{Merge Tables}  
This operation creates a new VBT from two or more VBTs. For each input table the  operation can use either all or a subset of the existing  columns. In case the VBTs do not have  the same number of rows, the operation creates a new VBT having the size of the smallest or of the larger VBT. In the first case longer columns will be truncated while in the second case the smallest columns are padded with user given values.

\subsubsection{Extract Subregions}  
This  operation creates a  new VBT by extracting a subregion
 from an input VBT using one of two geometric primitives, namely:
\begin{itemize}
\item {\it Sphere} - The center and the radius of the sphere determines the sub region to be extracted.
\item {\it Box} - The length of the box side and its corner position (or center position) determines the sub region to be extracted.
\end{itemize}
This operation can be used if the user needs to explore only a sub-volume from the original one.\\

\subsubsection{Select Rows}  
This operation creates a new VBT by setting limits on one or more fields of the input VBT. The operation must have a list of column names, and the limits (max-min) to define the extent of its action. Limits on more than one columns can be combined with OR or AND operators. The rows  that satisfy the given limits are reported on the output VBT. The output VBT contains all the columns of the input VBT.  For volumes, the Extract Subvolumes operation filter provides similar functonality (see 3.2.8).

\subsubsection{Mathematical Operations} 
These operations allow the user to derive new fields starting from the fields of an input VBT. As a result new columns are appended to the input VBT or (optionally) a new VBT is created. These operations are based on the function parser for C++ v2.83 by Warp \citep{Warp}  with some minor modifications.
As reported in the function parser library, a function string is very similar to the C-syntax. Arithmetic float expressions can be created from float literals, variables or functions using the usual mathematical operators and the standard mathematical functions.

\subsubsection{Point Distribution}  
This operation is used for distributing a scalar value throughout a regularly spaced 3D mesh adhering to a user prescribed resolution. The result is a volume computed by using a 3D Cloud-in-Cell (CIC) smoothing algorithm \citep{hockney}. The user can choose the column of an input VBT to be used for distributing and the 3D mesh resolution. 

\subsubsection{Point Property} 
This operation assigns a new property to each data point on a given VBT. The operation performs the following:
\begin{enumerate}
\item  A temporary  volume is created using a field distribution (CIC algorithm) on a regular mesh through the aforementioned {\it Point Distribution} operation.
\item Using a CIC algorithm again, the new property for each data point is computed taking into account all the 3-D cells  the point is associated with.
\item The property  is finally added as a new column into the input VBT, or a new VBT is created containing this new field only.
\end{enumerate}
This operation could have a typical application in case we want to assign a colour  to each point starting from any property. For example, if we want to assign a colour  depending on the mass density, we need to distribute the mass associated with points on a regular mesh. Then the colour  of each data point will be determined considering the nearest cell where each point is located.\\

\subsubsection{Extract Subvolumes} 
This  operation produces a  VBT  representing a subvolume of an existing VBT. The operation needs a starting cell and a number of cells along each coordinate direction. It can be used to investigate only a subvolume from the original one. All the columns of the input VBT are reported on the output VBT, and the analysis on all the volume fields of a given dataset could be done only on a specific region but with a reduced data size.

\subsubsection{Coarse Volumes}  
This  operation produces a coarse subvolume from an existing VBT using a plane extraction method. Planes are extracted uniformly along each of the coordinate directions. The operation needs the percentage of the input VBT the user wants to obtain. For example, if the original VBT is a mesh of 320x320x640 cells and the  percentage is $10\%$ the output VBT will be a mesh of 32x32x64 cells. This operation is a mandatory task for visualizing very large-scale datasets using VisIVO Viewer.

\subsubsection{Other Operations}  
VisIVO Filter also includes a range of other modules described briefly in this section. For more details please see the documentation found on \citep{visivoserver}. 
\begin{itemize}
\item {\it Append Tables} -  A  VBT is created appending two or more VBTs.
\item {\it Polar Transformation} -  This module performs a spherical polar transformation.
\item {\it Decimator} - A VBT is created as a regular (i.e. not randomized) subsample from an input VBT.
\item {\it Interpolate} - Intermediate VBTs are produced with a linear interpolation from two input VBTs.
\item {\it Show Table} - An ascii table from a VBT is printed.
\item {\it Sigma Contours} - A VBT is produced where  one or more fields of the input VBT have values within N sigma contours 
\item {\it Statistic} - The average, min and max value and the histogram of a VBT field is produced.
\item {\it Visual} - A new VBT  from  one or more input VBTs is created, appropriate for the visualization with VisIVO Viewer. All  input VBTs must have the same number of rows.
\end{itemize}
We are currently working in incorporating into VisIVO Filter a range of other modules, already available on VisIVO Desktop, such as calculation for correlation function, power spectrum and Minkowsky functionals.

\subsubsection{Performance Discussion}
\begin{figure}
\begin{center}
\includegraphics[width=6.5cm, angle=-90]{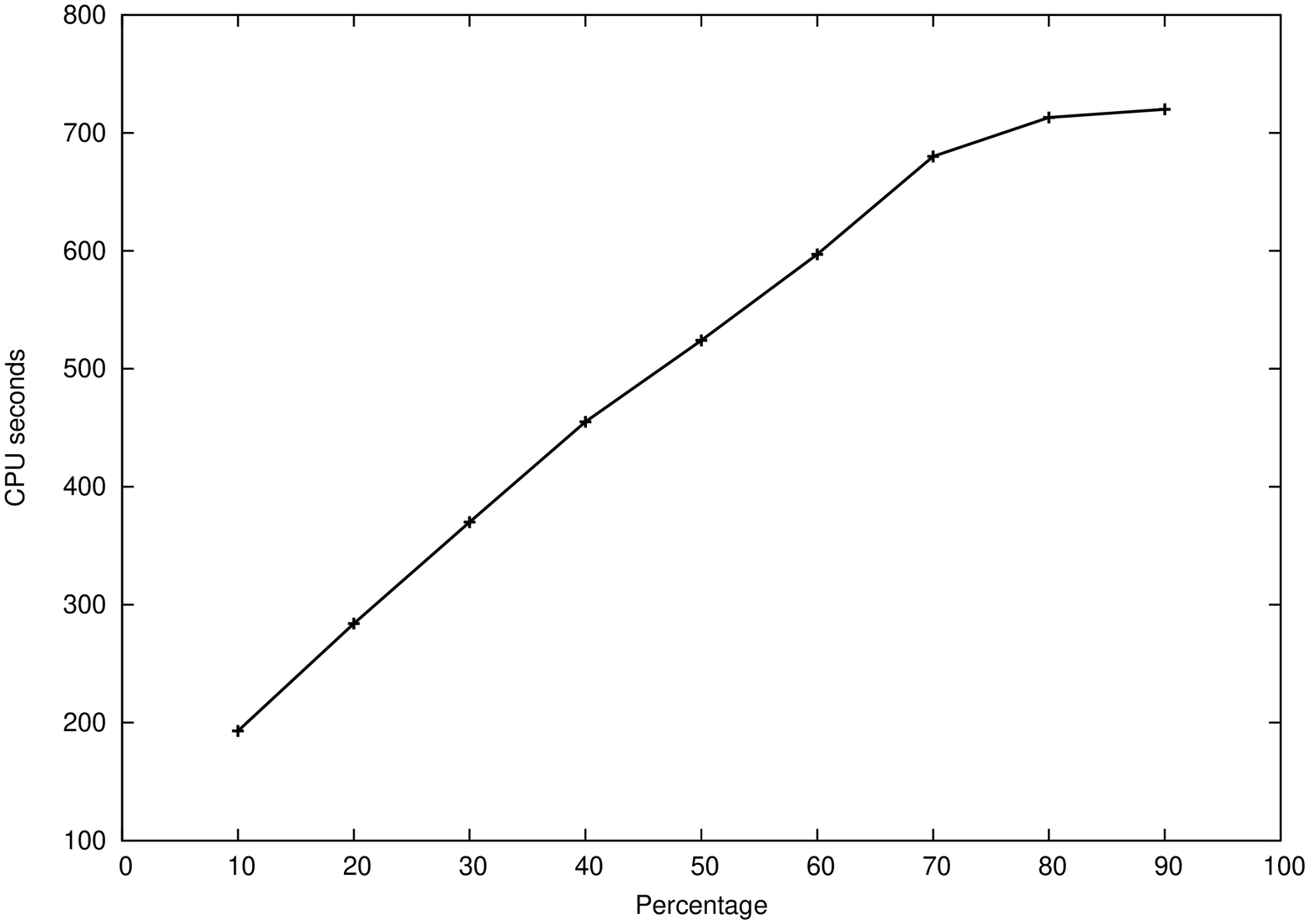}
\includegraphics[width=6.5cm, angle=-90]{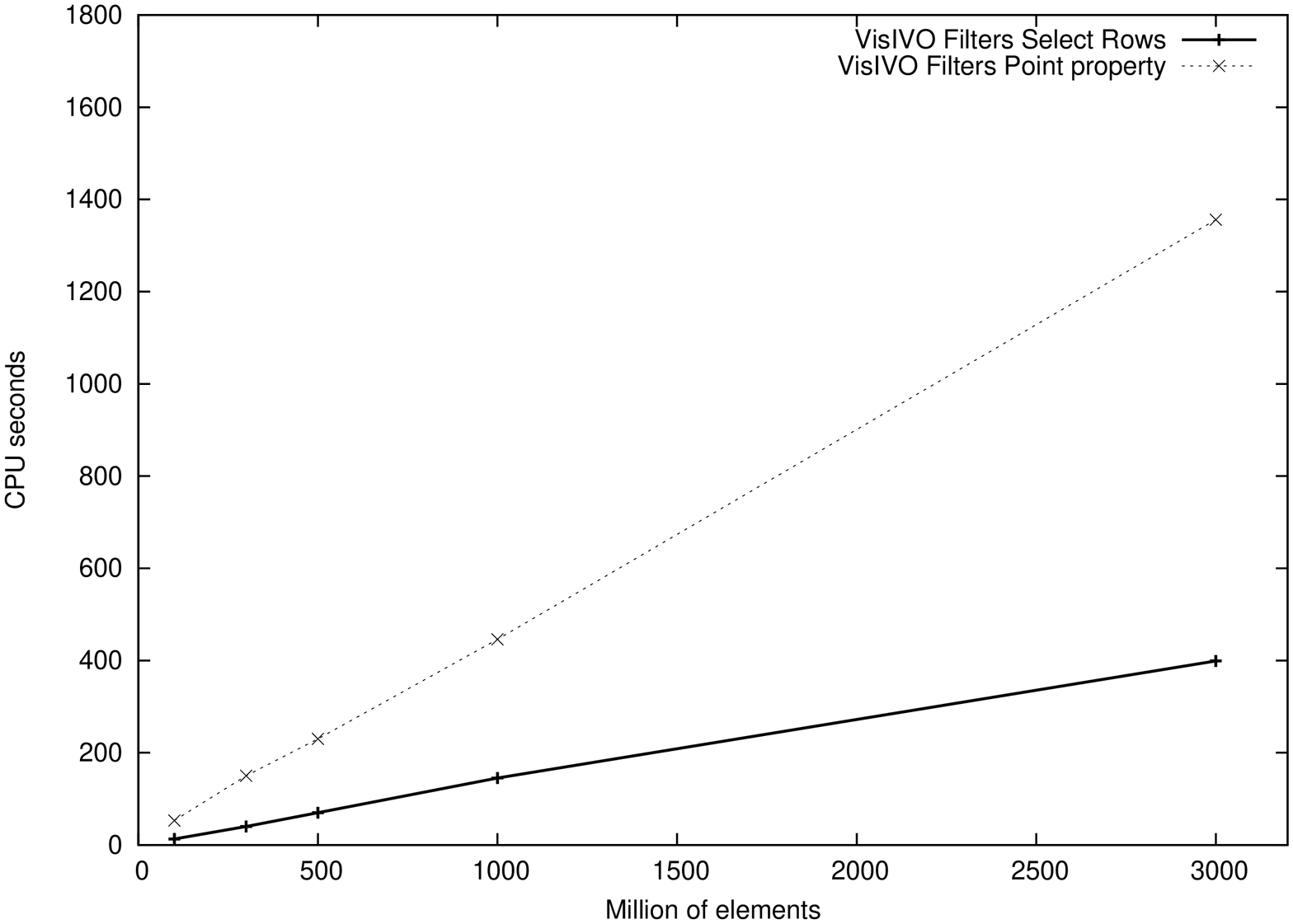}
\caption{VisIVO Filter Timings. Top: randomization operations for a Gadget simulation containing 3 billion elements. Bottom: select row and point property operations for different number of elements.}
\label{fil1}
\end{center}
\end{figure}

We tested the most common filters, namely {\it Randomization}, {\it Select Rows} and {\it Point Property} to assess overall performance timings. Figure \ref{fil1} shows the timings obtained with  datasets that can be fitted entirely within the available RAM but also for datasets that are larger thus requiring multiple disk read/write access. The randomization slope shown in the figures is determined by the number of read/write operations. The time to read the 3 billions input files is fixed, however the time to write the output files depends upon the percentage of the randomizer. The actual randomization process is however linear.\\
The figure also demonstrates the performance of the Select Row and Point Property filters. The performance of these operations is linear, a typical behaviour for all other filters in VisIVO Server. 

\subsection{VisIVO Viewer}
VisIVO Viewer is founded on the Visualization ToolKit version 5.2 \citep{vtk} library for multidimensional visualization using the Mesa library version 7.0.3 \citep{mesa} to avoid dependency upon an X Server connection. It creates 3D images of datasets, both data points and volumes can be represented. Though VisIVO Server can manage multidimensional datasets, the visualization process can be given on a large  number of elements. On the other hand, a visualization process with many million of elements (ten or more) does not typically give more significant visual information than some million of elements. Usually (with 2 GBRAM) VisIVO Viewer can display up to 16 million of data elements easily, but the visual information the user can have is not more rich 
than that given by two or four million of displayed elements.\\
The visualization process for large-scale datasets requires the following actions:
\begin{itemize}
\item VisIVO Filters running the {\it Randomization} operation. This operation reduces the original dataset to fit it into the available memory, and produces a new VBT for the visualization.
\item VisIVO Viewer running with the  VBT given by the aforementioned process. 
\end{itemize}
VisIVO Viewer can render points, volumes and isosurfaces within a bounding box used for representing the coordinate system employed. Moreover there is support for customized look up tables for visualization using a variety of glyphs, such as cubes, spheres or cones. The standard output of VisIVO Viewer consists of four images corresponding to fixed camera positions and zoom factors and another image corresponding to user-defined line command options.\\
The parameters for these images are as follows: zooming factor equals to $1$ and the Azimuth and Elevation of the camera are set to: 1) Az:0 - El: 0; 2) Az:90 - El: 0; 3) Az:0 - El: 90; 4) Az:45 - El: 45. The default images are always created even if the azimuth, elevation and zooming camera are assigned with user-defined values. They can be avoided with a {\it nodefault} option, prescribed by the user.\\
VisIVOViewer can be also used to produce images in a given sequence of azimuth elevation and zooming that can be externally mounted to produce a movie. Moreover VisIVO Viewer allows the visualization with Splotch \citep{Splotch} (see section 3.3.2).

\subsubsection{Data Points Views}
\begin{figure}
\begin{center}
\includegraphics[width=8.0cm]{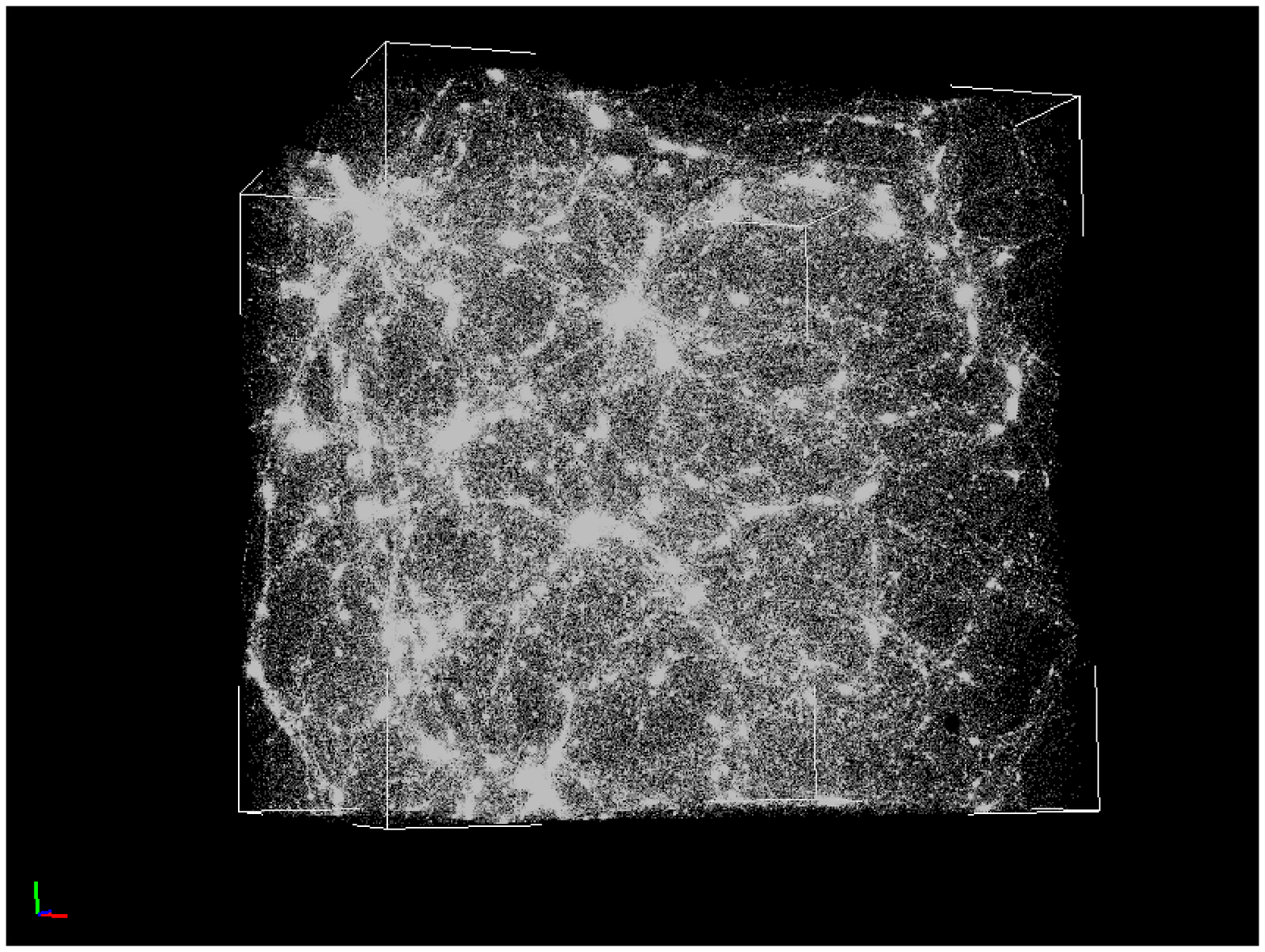}
\includegraphics[width=8.0cm]{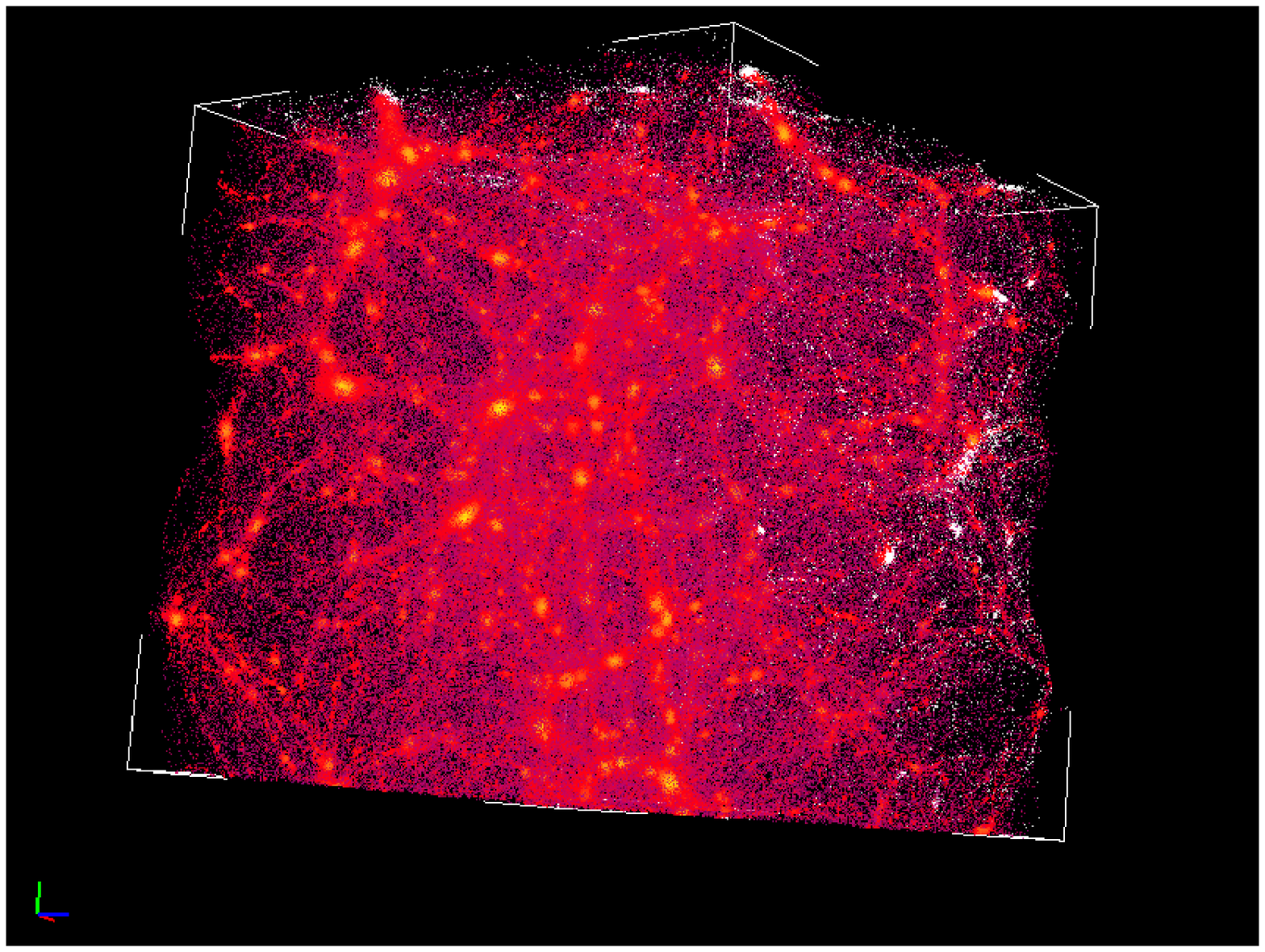}
\includegraphics[width=8.0cm]{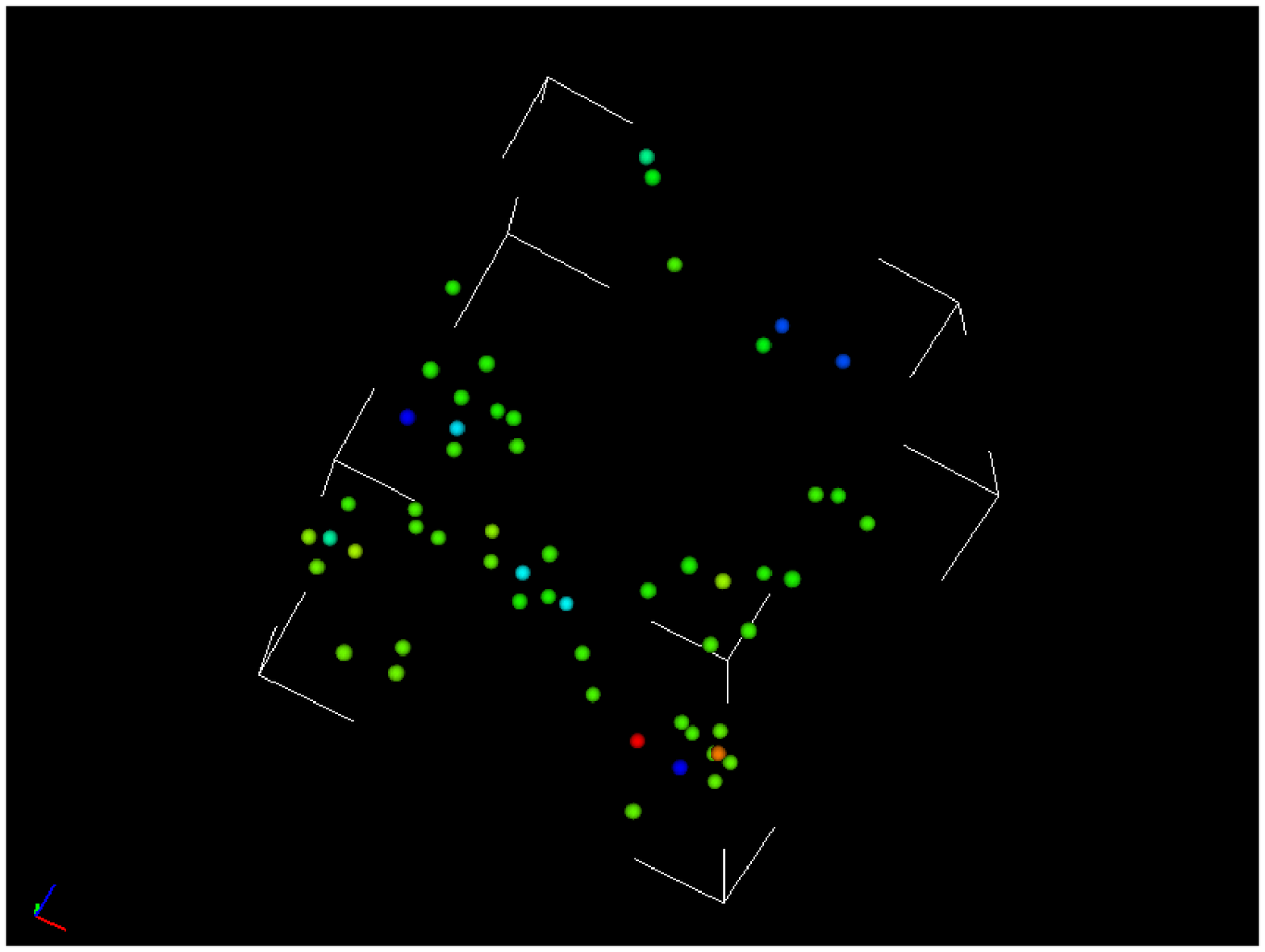}
\caption{VisIVO Viewer Data Points Views:  a cosmological N-body simulation containing 4 million elements;  the simulation elements are now coloured according to their mass density - opacity 0.3 is employed;  coloured glyph representation of 57 elements obtained from a Hypparcos catalogue query, M30 object with radius 40 arcsec, optical band}
\label{vie1}
\end{center}
\end{figure}

VisIVO Viewer uses any three columns of a VBT to create a 3D coordinate system where data  points can be drawn. If  the 3D coordinate system is not specified in the command options, it is created with columns having names starting with $X$, $Y$ and $Z$  or $Ra$, $De$ and $Mag$. Visualization of many points can give scientifically interesting images, e.g. consider looking for inner filaments and clusters, using an adequate opacity factor (Fig. \ref{vie1}).\\ 
If the axis scale values of the selected coordinate system are very different, the produced images could be very small and sometimes could even degenerate into a single point. To avoid this effect the {\it scale} option can be employed to always scale the coordinate axes into a cubic region.\\
Points can be coloured using a look-up (Lut) table among some pre-defined available look-up tables and using a logarithmic scale. A geometrical  form (or glyph) can be associated with each data point in the image: spheres, cones, cylinders and cubes can be used to represent each data point. This feature produces  good and very meaningful images when few elements must be displayed. Moreover the radius and the height of glyph elements can be scaled with two columns of the VBT. On the other hand, the glyphs visualization is  not allowed with an excessively  large number of elements. In fact the rendering process could last a very long time and very often the effect is a cubic region totally cluttered making  the produced images difficult to interpret.\\ 
VisIVO Viewer, with Lut and glyphs, can display up to six properties on the same image simultaneously, a very powerful feature for meaningful visualization and effective data exploration.
\subsubsection{Splotch Views}
\begin{figure}
\begin{center}
\includegraphics[width=7.5cm]{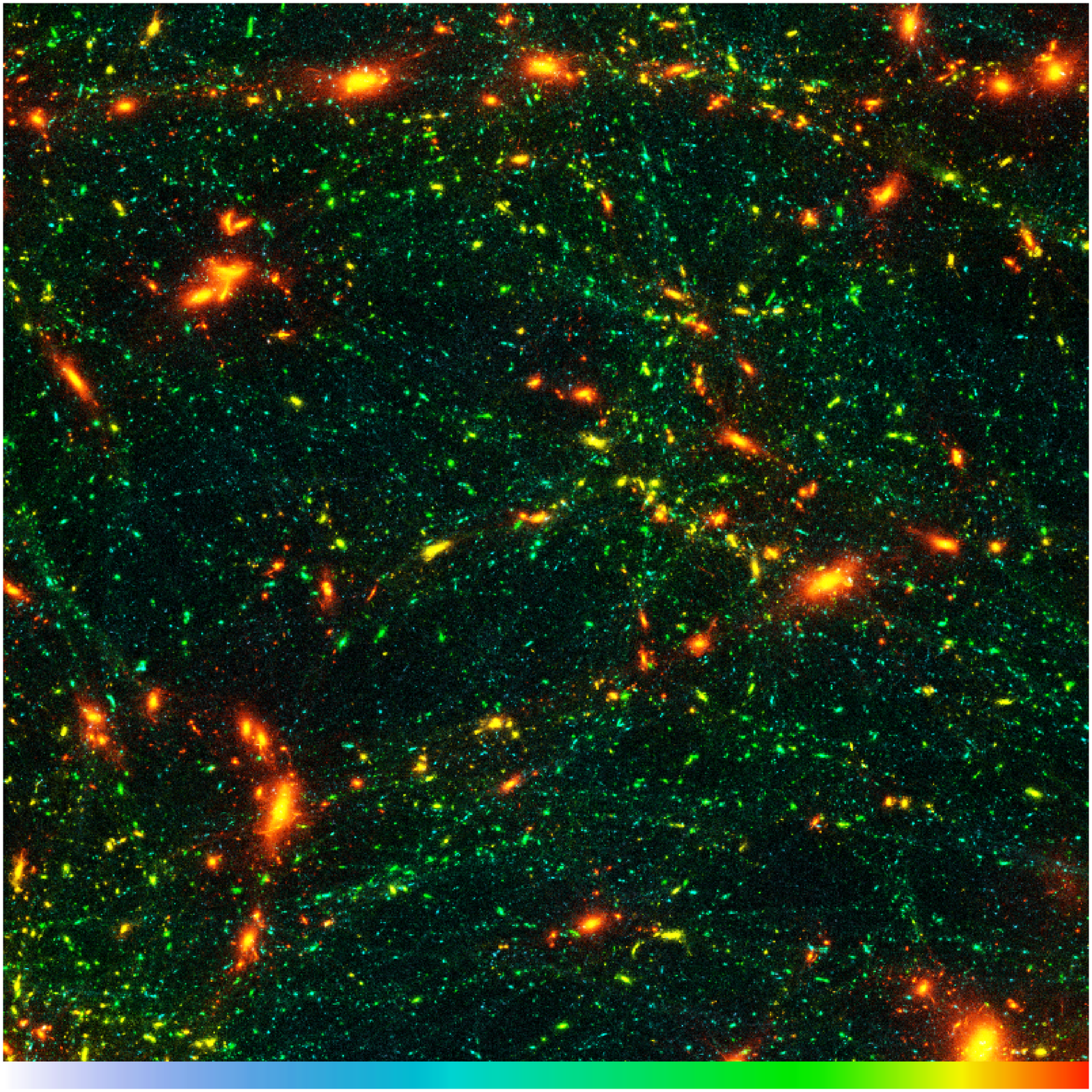}
\caption{Seventy million elements of a cosmological N-body simulation visualized with Splotch. Careful inspection reveals elliptical galaxies, dwarf ellipticals, merging galaxies and also filaments and voids.}
\label{spl1}
\end{center}
\end{figure}

Splotch  is a raytracer to visualize SPH simulations that we have recently customized to read data from a VBT. Splotch is designed  to deal with point-like data, optimizing the ray-tracing calculation by ordering the particles as a function of their {\it depth}, defined as a function of one of the coordinates or other associated parameters. Realistic three-dimensional renderings are reached through a composition of the final colour in each pixel by properly calculating emission and absorption of individual volume elements. Our customized version is included in the current VisIVO Server distribution. Many options used  to visualize Data Points can be used with Splotch (Fig. \ref{spl1}). The description of Splotch can be found on \citep{Splotch}. Currently we are working in optimising rendering times with splotch by following a hybrid parallelization approach exploiting multicore CPUs in conjunction with the functionality offered by modern underlying GPUs.

\subsubsection{Volume Views}
\begin{figure}
\begin{center}
\includegraphics[width=7.0cm]{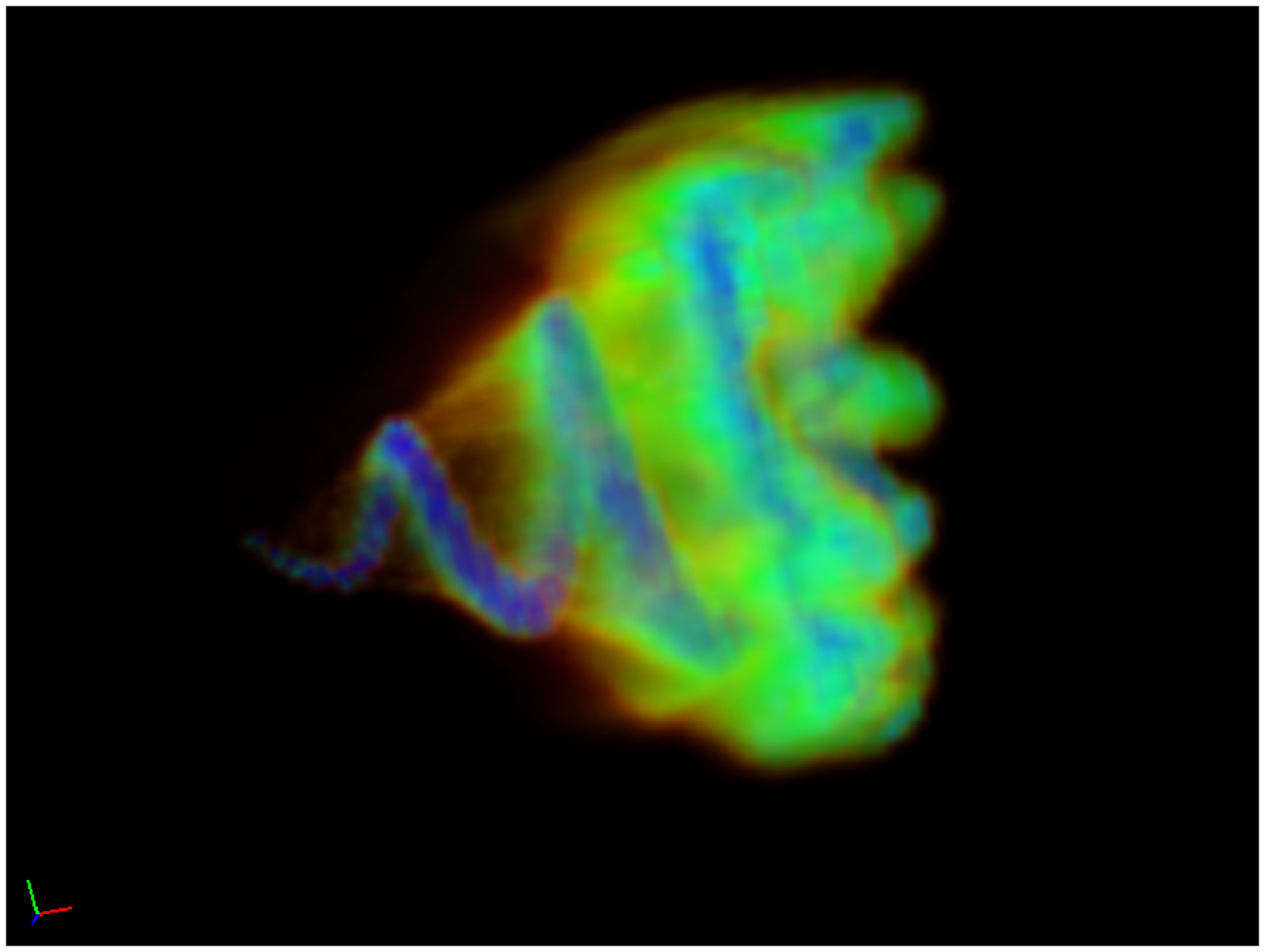}
\includegraphics[width=9.0cm]{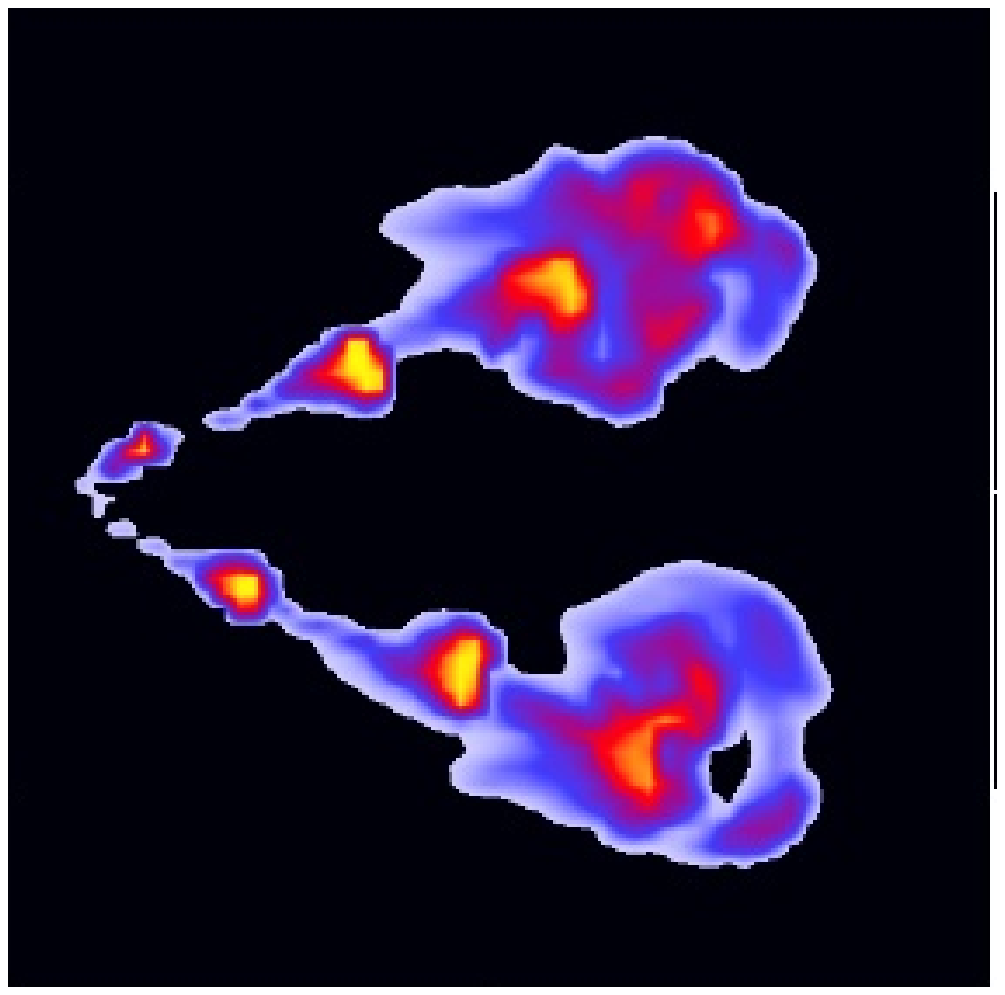}
\includegraphics[width=7.0cm]{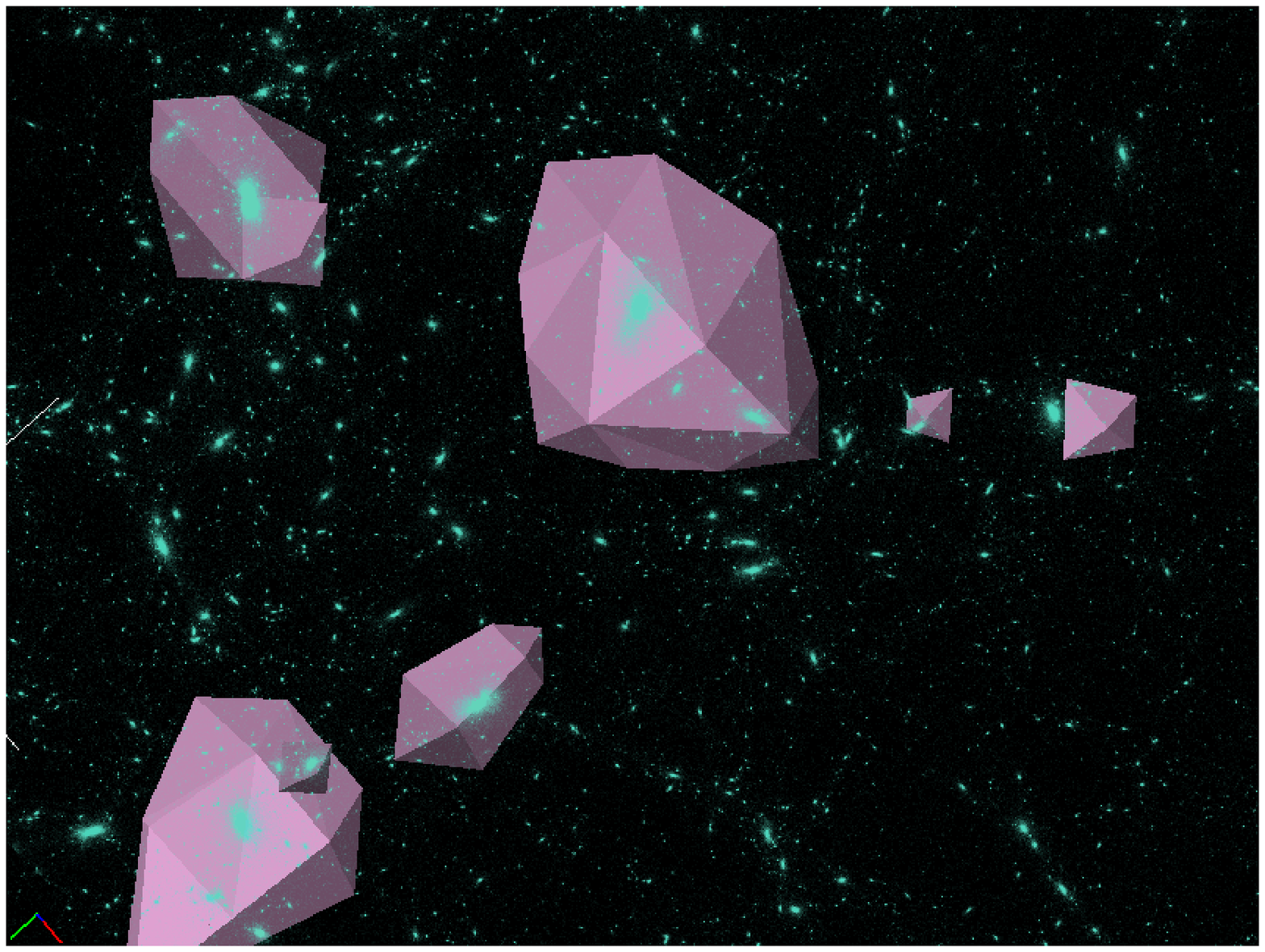}
\caption{VisIVO Viewer Volume Views:  volume rendering of a relativistic jet, i.e. a relativistic electron-positron plasma flow emitted by an active galactic nucleus containing a black hole;  the 3D slicer is used for inspection of cross-sections through the relativistic jet;  isosurfaces of galaxy clusters formed in an N-body cosmological simulation - these could be subsequently used for extracting the relevant part of the simulation for further analysis.}
\label{vie2}
\end{center}
\end{figure}

VisIVO Viewer includes algorithms for direct volume rendering and iso-surfacing (Fig. \ref{vie2}, top and bottom) for regular 3D meshes created from a point distribution operation (see 3.2.6). The underlying rendering functionality is founded on the Visualization Toolkit. However for large-scale datasets the standard iso-surfacing algorithm execution is slow. Further as the resulting surface models generally contain very large numbers of tiny triangles, the overall computational performance can be degraded considerably, e.g. in generating sequences of renderings for producing movies. We have previously implemented optimised rendering algorithms in VisIVO Desktop and we are currently porting these into VisIVO viewer. The iso-surfacing algorithm is optimised significantly in several ways by using multiresolution, min-max blocks, point caching and finally multi-threading. Multiresolution allows us to generate lightweight preview models. Min-max blocks pre-compute minimum and maximum values for data blocks, thus allowing to discard irrelevant data blocks (that is, with values below or above the iso-surface threshold) during rendering very quickly. Finally, point caching ensures that no extra computations are performed for polygonal model vertices. The direct volume rendering algorithm employs similar optimisations by using multi-resolution data sampling, adaptive pixel sampling and render caching, thus allowing fast algorithm execution with high rendering quality. Colouring is implemented through user-prescribed look up tables which are selected from a library of several pre-defined look-up tables. VisIVO Viewer also provides a computationally efficient 3D slicer for visualization of orthogonal cross-sections cutting through regular 3D meshes (Fig. \ref{vie2}, middle).

\subsubsection{VisIVO Viewer Performance}
VisIVO Viewer implements a range of rendering algorithms, e.g. for points, isosurfaces and volumes, supporting high-performance visualization. For example, the speed improvement of our optimised iso-surfacing compared to standard VTK iso-surfacing is normally in the range of 30-100 times faster (even without multiresolution). This performance gives users the ability to select an optimal isosurface threshold interactively, an impossible task with the standard implementation.\\
The rendering performance for standard astrophysical datasets (i.e. datasets typically fitting into the local PC's memory) is very satisfactory for our purposes, e.g. no more than a few seconds are often required to generate renderings.\\
Nevertheless rendering using the Splotch algorithm could involve substantial computational costs, e.g. several minutes or even longer times depending upon the values assigned to the Splotch rendering parameters. Our experimental results indicate that the value chosen for smoothing length exhibits a strong influence on overall rendering times. Typically the user needs to do some experimentation to strike a balance when choosing an appropriate smoothing length. Too large values may result in prohibitively excessive computational costs. On the other hand, too small values may result in flat renderings i.e. without fog-like effects. 

\section {VisIVO on the Grid}
The development of the VisIVO family visualization tools was initiated within the EU funded project VO-TECH \citep{VOTech}. The research collaborations with the Cometa Consortium in Catania, Italy, and the University of Portsmouth, United Kingdom, produced VisIVO Server. To include the possibility of handling large-scale computational tasks we have recently implemented a grid version of VisIVO Server.\\ Our current implementation exploits the infrastructure of the Cometa Consortium with main grid nodes located in the Sicilian cities of Catania, Messina and Palermo respectively. The underlying hardware is based on IBM Blade Centre technology each containing up to 14 IBM LS21 blades interconnected with a low latency Infiniband-4X network. Further, each blade is equipped with 2 AMD Opteron 2218 rev. F dual-core processors with a clock rate of 2,6 GHz allowing to native execute x86 instructions in 32 and 64 bits. Overall the currently employed grid infrastructure  consists of over 2,500 CPU cores \citep{ube} \\ The grid version of VisIVO Server is a very powerful environment offering data analysis and exploration of very large-scale astrophysical datasets. For example during a lengthy run for generating a highly complex numerical simulation, the system allows users to analyse and visually explore the simulation - in other words it provides a way to monitor the simulation as it happens. A forthcoming article will discuss in detail our experiences with VisIVO Server in grid environments.

\section{VisIVO Web}
\begin{figure}
\begin{center}
\includegraphics[width=9.0cm]{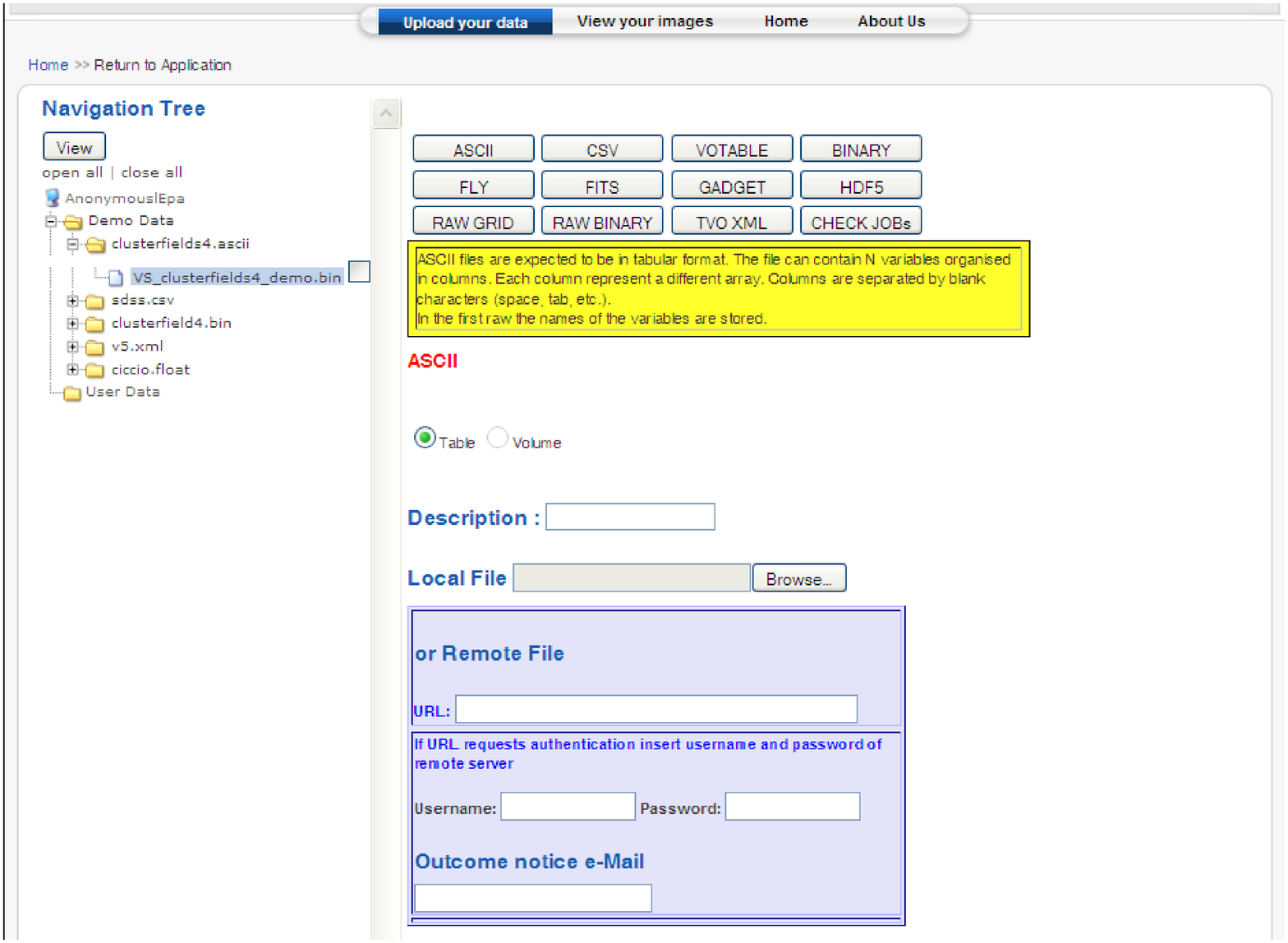}
\includegraphics[width=12.0cm]{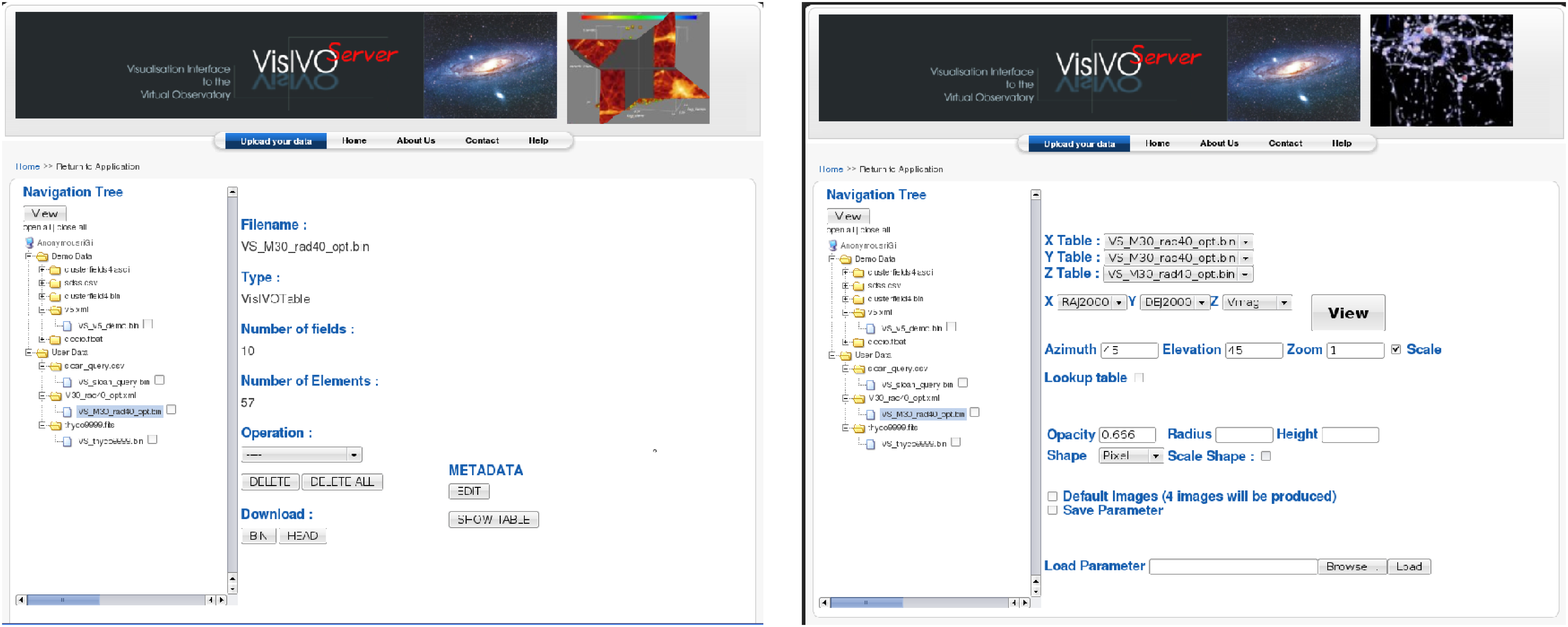}
\includegraphics[width=8.0cm]{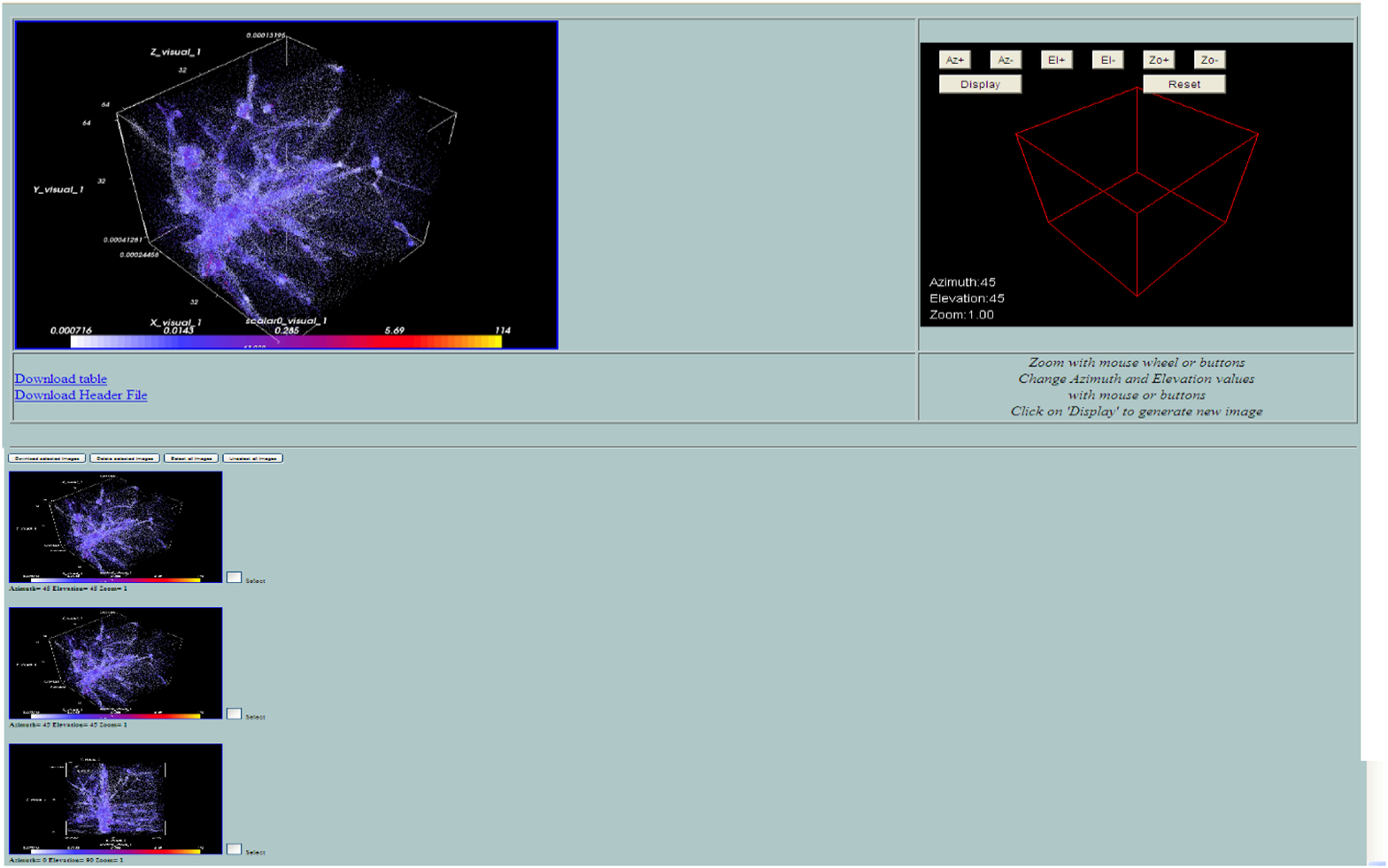}
\caption{VisIVO Web Operational Scenario. The top figure shows uploading of datasets from a user's PC or from a server remotely. The middle figures demonstrate the application of VisIVO Filter operations. The bottom figure illustrates the final display using customized colour tables and interactive widgets for setting VisIVO Viewer camera positions. A detailed description of the VisIVO Web operational scenario can be found in the user guide which can be downloaded from http://visivoserver.oact.inaf.it.}
\label{web}
\end{center}
\end{figure}

VisIVO Web is a recently developed www portal providing VisIVO Server services to the scientific community. Our purpose was to offer prospective users an intuitive and easy-to-use graphical environment for accessing the full functionality of VisIVO Server. This section gives an overview of VisIVO Web, a detailed presentation of its functionality will be given in a forthcoming publication (Fig. \ref{web}) .\\ Users can upload and manage their datasets (by using registered or anonymous access), e.g. by using interactive widgets to construct customized renderings, or storing data analysis and visualization results for future reference. The datasets are managed internally through a relational database for preserving any metadata and maintaining data consistency. Both remote and local datasets can be uploaded - i.e. residing on a remote www address or locally on a user's PC. For remote files the user must specify the relevant www address and optionally a username and password for authentication. Depending upon the size of the datasets under consideration, remote uploads could last a long period. To resolve this situation VisIVO Web allows an off-line mode so that users can issue upload commands and then simply close their current session - a follow up e-mail typically gives notification once the uploading operation is completed.\\
Once data is uploaded a sequence of simple actions is required to rapidly obtain meaningful visualizations. Typically various VisIVO Filter operations are performed, and VisIVO Web automatically displays all applicable VisIVO Filter operations allowing for graphical input of the relevant parameters. Finally the VisIVO Server Viewer is employed for display. A check box located on the right hand side of any processed dataset is used in conjunction with the View button to create user-prescribed VisIVO Viewer views. VisIVO Web is currently realised in www sites hosted by the University of Portsmouth, UK, the INAF Astrophysical Observatory of Catania and Astronomical Observatory of Trieste, Italy \citep{visivoserver}.

\section{Installation Notes}
VisIVO Server is an open source software available to the scientific community on SourceForge that provides free hosting to Open Source software development projects \citep{sourceforge}. VisIVO Server requires as a minimum the following libraries: gcc 4.3, cmake  2.6, xerces 2.8, cfitsio 3.1, curl 7.19.2, vtk 5.2 and mesa 7.0.3. These libraries must be installed on the underlying server and their compilation can be done either dynamically or statically. Standard installation of VisIVO Server does not require an X Server, so that no further restrictions are imposed (e.g. firewalls do not typically allow to open X Windows remotely).

\section{Summary}
VisIVO is an integrated suite of tools and services specifically designed for the Virtual Observatory. This paper focused on VisIVO Server, a new platform for astrophysical visualization of large-scale datasets that can be easily installed on any computing server. VisIVO Server is an open-source collection of data processing and visualisation modules allowing fast rendering of 3D views. VisIVO Server is founded on the functionality of VisIVO Desktop, our previously developed stand-alone astrophysical visualization application for standard PCs.The defining characteristic of VisIVO Server is that no restrictions are imposed on dimensionality of datasets.\\ We described the constituent parts of VisIVO Server in detail, namely VisIVO Importer, VisIVO Filter and VisIVO Viewer. VisiVO Importer converts user prescribed datasets into a highly efficient internal data format employed by VisIVO Filters and VisIVO Viewer. VisIVO Filters are a collection of several processing modules for constructing customised data tables from VisIVO Importer. Finally, VisIVO Viewer creates 3D views of astrophysical datasets, currently supporting points, volumes iso-surfaces and ray-tracing using a customized algorithm called Splotch.\\ We discussed different visualization scenarios and demonstrated example renderings. Users can obtain meaningful visualizations rapidly while preserving full and intuitive control of the relevant visualization parameters. We also described briefly a computational grid realisation of VisIVO Server and also discussed VisIVO Web - a custom designed web portal supporting services based on the VisIVO Server functionality. The VisIVO Server software is distributed under a GPL license  for non-commercial use only. It is an open source project and can be simply downloaded from the sourceforge code repository.\\
VisIVO Server and VisIVO Desktop can exchange data using the same internal data format (VBT). They can be closely integrated, but are also complementary and independent of each other. For example VisIVO Server can create a preview of a dataset or an appropriately-defined subsampled version, then VIsIVO Desktop can be employed for visualization and subsequent interaction.\\ 
Currently we are working towards an integrated visualisation functionality so that VisIVO Server and VisIVO Desktop are wrapped around a unified visualisation kernel. Our vision is to maintain VisIVO Server for high performance computing (e.g. grid environments, large memories) and VisIVO Desktop to support interactive front-end interfaces providing an advanced GUI and exploiting the full graphical capabilities of modern PCs (e.g. fast multi-processor architectures, emerging GPUs). Towards this aim we will need to define a full communication protocol between the Server and Desktop versions, so that applications for each will be possible to be developed independently.

\section{Acknowledgements}
The authors would like to acknowledge the support of Dr. Roberto Munzone for his significant contribution during the development of the Gadget reader, and Dr. P. Manzato and Dr. M. Molinaro from the INAF Astronomical Observatory of Trieste for several useful discussions during the course of this project. Our special thanks to F. Vitello from the Cometa Consortium for developing parts of the VisIVO website. Finally we would like to thank Prof. B. Nichol from the Institute of Cosmology and Gravitation, University of Portsmouth for his support during the development of VisIVO Server. This work has made use of results produced by the PI2S2 Project managed by the Consorzio COMETA, a project co-funded by the Italian Ministero dell'Istruzione, Universita' e Ricerca (MIUR) within the Piano Operativo Nazionale Ricerca Scientifica, Sviluppo Tecnologico, Alta Formazione (PON 2000-2006). More information is available at 
http://www.pi2s2.it and http://www.consorzio-cometa.it. The work reported in this paper was partially supported by the HPC-EUROPA++ (project number 211437), under the EC Coordination and Support Action Research Infrastructure Programme in FP7 and by the University of Portsmouth, promising researcher’s award. 
\bibliographystyle{elsart-harv}
\bibliography{biblio}

\begin{thebibliography}{30}
\expandafter\ifx\csname natexlab\endcsname\relax\def\natexlab#1{#1}\fi
\expandafter\ifx\csname url\endcsname\relax
  \def\url#1{\texttt{#1}}\fi
\expandafter\ifx\csname urlprefix\endcsname\relax\def\urlprefix{URL }\fi

\bibitem[{{Becciani}(2007)}]{ube}
{Becciani}, U., 2007. {New Grid Infrastructure in Sicily}. Computational Grids
  for Italian Astrophysics: Status and Perspectives Ed. L. Benacchio, F.
  Pasian, 179--189.

\bibitem[{{Becciani}(2009)}]{BeccianiGrid}
{Becciani}, U., 2009. {The Cometa Consortium and the PI2S2 project .} Memorie
  della Societa Astronomica Italiana Supplement 13, 10--+.

\bibitem[{{Becciani} et~al.(2007){Becciani}, {Antonuccio-Delogu}, and
  {Comparato}}]{2007CoPhC.176..211B}
{Becciani}, U., {Antonuccio-Delogu}, V., {Comparato}, M., Feb. 2007. {FLY:
  MPI-2 high resolution code for LSS cosmological simulations}. Computer
  Physics Communications 176, 211--217.

\bibitem[{{Bonnarel} et~al.(2000){Bonnarel}, {Fernique}, {Bienaym{\'e}},
  {Egret}, {Genova}, {Louys}, {Ochsenbein}, {Wenger}, and {Bartlett}}]{Aladin}
{Bonnarel}, F., {Fernique}, P., {Bienaym{\'e}}, O., {Egret}, D., {Genova}, F.,
  {Louys}, M., {Ochsenbein}, F., {Wenger}, M., {Bartlett}, J.~G., Apr. 2000.
  {The ALADIN interactive sky atlas. A reference tool for identification of
  astronomical sources}. \aaps 143, 33--40.

\bibitem[{{Comparato} et~al.(2007){Comparato}, {Becciani}, {Costa}, {Larsson},
  {Garilli}, {Gheller}, and {Taylor}}]{visivo}
{Comparato}, M., {Becciani}, U., {Costa}, A., {Larsson}, B., {Garilli}, B.,
  {Gheller}, C., {Taylor}, J., Aug. 2007. {Visualization, Exploration, and Data
  Analysis of Complex Astrophysical Data}. \pasp 119, 898--913.

\bibitem[{{Costa} et~al.(2008){Costa}, {Manzato}, {Becciani}, {Comparato},
  {Costa}, {Gasparo}, {Gheller}, {Grillo}, {Molinaro}, {Pasian}, and
  {Taffoni}}]{costa}
{Costa}, A., {Manzato}, P., {Becciani}, U., {Comparato}, M., {Costa}, V.,
  {Gasparo}, F., {Gheller}, C., {Grillo}, A., {Molinaro}, M., {Pasian}, F.,
  {Taffoni}, G., Aug. 2008. {The TVO Archive for Cosmological Simulations: Web
  Services and Architecture}. \pasp 120, 933--944.

\bibitem[{{Dolag} et~al.(2008){Dolag}, {Reinecke}, {Gheller}, and
  {Imboden}}]{Splotch}
{Dolag}, K., {Reinecke}, M., {Gheller}, C., {Imboden}, S., Dec. 2008. {Splotch:
  visualizing cosmological simulations}. New Journal of Physics 10~(12),
  125006--+.

\bibitem[{{Hockney} and {Eastwood}(1988)}]{hockney}
{Hockney}, R.~W., {Eastwood}, J.~W., 1988. {Computer simulation using
  particles}.

\bibitem[{http://curl.haxx.se(2010)}]{Curl}
http://curl.haxx.se, 2010.

\bibitem[{http://eurovotech.org(2010)}]{VOTech}
http://eurovotech.org, 2010.

\bibitem[{http://fits.gsfc.nasa.gov/(2010)}]{Fits}
http://fits.gsfc.nasa.gov/, 2010.

\bibitem[{http://iki.fi/warp/FunctionParser/(2010)}]{Warp}
http://iki.fi/warp/FunctionParser/, 2010.

\bibitem[{http://sourceforge.net/(2010)}]{sourceforge}
http://sourceforge.net/, 2010.

\bibitem[{https://wci.llnl.gov/codes/visit/(2010)}]{wwwVisit}
https://wci.llnl.gov/codes/visit/, 2010.

\bibitem[{http://visivo.oact.inaf.it/(2010)}]{wwwvisivodk}
http://visivo.oact.inaf.it/, 2010.

\bibitem[{http://visivoserver.oact.inaf.it(2010)}]{visivoserver}
http://visivoserver.oact.inaf.it, 2010.

\bibitem[{http://www.ivoa.net/cgi
  bin/twiki/bin/view/IVOA/IvoaVOTable(2010)}]{VOTable}
http://www.ivoa.net/cgi bin/twiki/bin/view/IVOA/IvoaVOTable, 2010.

\bibitem[{http://www.lofar.org/(2010)}]{LOFAR}
http://www.lofar.org/, 2010.

\bibitem[{http://www.mesa3d.org/(2010)}]{mesa}
http://www.mesa3d.org/, 2010.

\bibitem[{http://www.sdss.org/(2010)}]{WWWSDSS}
http://www.sdss.org/, 2010.

\bibitem[{http://www.slicer.org/(2010)}]{wwwSlicer}
http://www.slicer.org/, 2010.

\bibitem[{http://www.starlink.ac.uk/topcat/(2010)}]{Topcat}
http://www.starlink.ac.uk/topcat/, 2010.

\bibitem[{http://www.vtk.org/(2010)}]{vtk}
http://www.vtk.org/, 2010.

\bibitem[{{Ivezic} et~al.(2008){Ivezic}, {Axelrod}, {Brandt}, {Burke},
  {Claver}, {Connolly}, {Cook}, {Gee}, {Gilmore}, {Jacoby}, {Jones}, {Kahn},
  {Kantor}, {v.~Krabbendam}, {Lupton}, {Monet}, {Pinto}, {Saha}, {Schalk},
  {Schneider}, {Strauss}, {Stubbs}, {Sweeney}, {Szalay}, {Thaler}, and
  {Tyson}}]{LSST}
{Ivezic}, Z., {Axelrod}, T., {Brandt}, W.~N., {Burke}, D.~L., {Claver}, C.~F.,
  {Connolly}, A., {Cook}, K.~H., {Gee}, P., {Gilmore}, D.~K., {Jacoby}, S.~H.,
  {Jones}, R.~L., {Kahn}, S.~M., {Kantor}, J.~P., {v.~Krabbendam}, V.,
  {Lupton}, R.~H., {Monet}, D.~G., {Pinto}, P.~A., {Saha}, A., {Schalk}, T.~L.,
  {Schneider}, D.~P., {Strauss}, M.~A., {Stubbs}, C.~W., {Sweeney}, D.,
  {Szalay}, A., {Thaler}, J.~J., {Tyson}, J.~A., Jun. 2008. {Large Synoptic
  Survey Telescope: From Science Drivers To Reference Design}. Serbian
  Astronomical Journal 176, 1--13.

\bibitem[{{Mohr} and {Adams}~D.(2008)}]{DES}
{Mohr}, J., {Adams}~D., e.~a., Jul. 2008. {The Dark Energy Survey data
  management system}. In: Data Management and Quality Control. Vol. 7016 of
  SPIE Proceeedings. p. 70160L.

\bibitem[{{Price}(2007)}]{Splash}
{Price}, D.~J., Oct. 2007. {splash: An Interactive Visualisation Tool for
  Smoothed Particle Hydrodynamics Simulations}. Publications of the
  Astronomical Society of Australia 24, 159--173.

\bibitem[{{Schilizzi} et~al.(2008){Schilizzi}, {Dewdney}, and {Lazio}}]{SKA}
{Schilizzi}, R.~T., {Dewdney}, P.~E.~F., {Lazio}, T.~J.~W., Aug. 2008. {The
  Square Kilometre Array}. In: Society of Photo-Optical Instrumentation
  Engineers (SPIE) Conference Series. Vol. 7012 of Presented at the Society of
  Photo-Optical Instrumentation Engineers (SPIE) Conference.

\bibitem[{{Springel}(2005)}]{Gadget}
{Springel}, V., Dec. 2005. {The cosmological simulation code GADGET-2}. \mnras
  364, 1105--1134.

\bibitem[{{Viceconti} et~al.(2004){Viceconti}, {Astolfi}, {Leardini},
  {Imboden}, {Petrone}, {Quadrani}, {Taddei}, {Testi}, and {Zannoni}}]{maf}
{Viceconti}, M., {Astolfi}, L., {Leardini}, A., {Imboden}, S., {Petrone}, M.,
  {Quadrani}, P., {Taddei}, F., {Testi}, D., {Zannoni}, C., Jul. 2004. {The
  Multimod Application Framework}. In: Information Visualisation, 2004. IV
  2004. Proceedings. Eighth International Conference on. IEEE Proceeedings.

\bibitem[{{York} et~al.(2000){York}, {Adelman}, {Anderson}, {Anderson},
  {Annis}, {Bahcall}, {Bakken}, {Barkhouser}, {Bastian}, {Berman}, {Boroski},
  {Bracker}, {Briegel}, {Briggs}, {Brinkmann}, {Brunner}, {Burles}, {Carey},
  {Carr}, {Castander}, {Chen}, {Colestock}, {Connolly}, {Crocker}, {Csabai},
  {Czarapata}, {Davis}, {Doi}, {Dombeck}, {Eisenstein}, {Ellman}, {Elms},
  {Evans}, {Fan}, {Federwitz}, {Fiscelli}, {Friedman}, {Frieman}, {Fukugita},
  {Gillespie}, {Gunn}, {Gurbani}, {de Haas}, {Haldeman}, {Harris}, {Hayes},
  {Heckman}, {Hennessy}, {Hindsley}, {Holm}, {Holmgren}, {Huang}, {Hull},
  {Husby}, {Ichikawa}, {Ichikawa}, {Ivezi{\'c}}, {Kent}, {Kim}, {Kinney},
  {Klaene}, {Kleinman}, {Kleinman}, {Knapp}, {Korienek}, {Kron}, {Kunszt},
  {Lamb}, {Lee}, {Leger}, {Limmongkol}, {Lindenmeyer}, {Long}, {Loomis},
  {Loveday}, {Lucinio}, {Lupton}, {MacKinnon}, {Mannery}, {Mantsch}, {Margon},
  {McGehee}, {McKay}, {Meiksin}, {Merelli}, {Monet}, {Munn}, {Narayanan},
  {Nash}, {Neilsen}, {Neswold}, {Newberg}, {Nichol}, {Nicinski}, {Nonino},
  {Okada}, {Okamura}, {Ostriker}, {Owen}, {Pauls}, {Peoples}, {Peterson},
  {Petravick}, {Pier}, {Pope}, {Pordes}, {Prosapio}, {Rechenmacher}, {Quinn},
  {Richards}, {Richmond}, {Rivetta}, {Rockosi}, {Ruthmansdorfer}, {Sandford},
  {Schlegel}, {Schneider}, {Sekiguchi}, {Sergey}, {Shimasaku}, {Siegmund},
  {Smee}, {Smith}, {Snedden}, {Stone}, {Stoughton}, {Strauss}, {Stubbs},
  {SubbaRao}, {Szalay}, {Szapudi}, {Szokoly}, {Thakar}, {Tremonti}, {Tucker},
  {Uomoto}, {Vanden Berk}, {Vogeley}, {Waddell}, {Wang}, {Watanabe},
  {Weinberg}, {Yanny}, and {Yasuda}}]{SDSS}
{York}, D.~G., {Adelman}, J., {Anderson}, Jr., J.~E., {Anderson}, S.~F.,
  {Annis}, J., {Bahcall}, N.~A., {Bakken}, J.~A., {Barkhouser}, R., {Bastian},
  S., {Berman}, E., {Boroski}, W.~N., {Bracker}, S., {Briegel}, C., {Briggs},
  J.~W., {Brinkmann}, J., {Brunner}, R., {Burles}, S., {Carey}, L., {Carr},
  M.~A., {Castander}, F.~J., {Chen}, B., {Colestock}, P.~L., {Connolly}, A.~J.,
  {Crocker}, J.~H., {Csabai}, I., {Czarapata}, P.~C., {Davis}, J.~E., {Doi},
  M., {Dombeck}, T., {Eisenstein}, D., {Ellman}, N., {Elms}, B.~R., {Evans},
  M.~L., {Fan}, X., {Federwitz}, G.~R., {Fiscelli}, L., {Friedman}, S.,
  {Frieman}, J.~A., {Fukugita}, M., {Gillespie}, B., {Gunn}, J.~E., {Gurbani},
  V.~K., {de Haas}, E., {Haldeman}, M., {Harris}, F.~H., {Hayes}, J.,
  {Heckman}, T.~M., {Hennessy}, G.~S., {Hindsley}, R.~B., {Holm}, S.,
  {Holmgren}, D.~J., {Huang}, C., {Hull}, C., {Husby}, D., {Ichikawa}, S.,
  {Ichikawa}, T., {Ivezi{\'c}}, {\v Z}., {Kent}, S., {Kim}, R.~S.~J., {Kinney},
  E., {Klaene}, M., {Kleinman}, A.~N., {Kleinman}, S., {Knapp}, G.~R.,
  {Korienek}, J., {Kron}, R.~G., {Kunszt}, P.~Z., {Lamb}, D.~Q., {Lee}, B.,
  {Leger}, R.~F., {Limmongkol}, S., {Lindenmeyer}, C., {Long}, D.~C., {Loomis},
  C., {Loveday}, J., {Lucinio}, R., {Lupton}, R.~H., {MacKinnon}, B.,
  {Mannery}, E.~J., {Mantsch}, P.~M., {Margon}, B., {McGehee}, P., {McKay},
  T.~A., {Meiksin}, A., {Merelli}, A., {Monet}, D.~G., {Munn}, J.~A.,
  {Narayanan}, V.~K., {Nash}, T., {Neilsen}, E., {Neswold}, R., {Newberg},
  H.~J., {Nichol}, R.~C., {Nicinski}, T., {Nonino}, M., {Okada}, N., {Okamura},
  S., {Ostriker}, J.~P., {Owen}, R., {Pauls}, A.~G., {Peoples}, J., {Peterson},
  R.~L., {Petravick}, D., {Pier}, J.~R., {Pope}, A., {Pordes}, R., {Prosapio},
  A., {Rechenmacher}, R., {Quinn}, T.~R., {Richards}, G.~T., {Richmond}, M.~W.,
  {Rivetta}, C.~H., {Rockosi}, C.~M., {Ruthmansdorfer}, K., {Sandford}, D.,
  {Schlegel}, D.~J., {Schneider}, D.~P., {Sekiguchi}, M., {Sergey}, G.,
  {Shimasaku}, K., {Siegmund}, W.~A., {Smee}, S., {Smith}, J.~A., {Snedden},
  S., {Stone}, R., {Stoughton}, C., {Strauss}, M.~A., {Stubbs}, C., {SubbaRao},
  M., {Szalay}, A.~S., {Szapudi}, I., {Szokoly}, G.~P., {Thakar}, A.~R.,
  {Tremonti}, C., {Tucker}, D.~L., {Uomoto}, A., {Vanden Berk}, D., {Vogeley},
  M.~S., {Waddell}, P., {Wang}, S., {Watanabe}, M., {Weinberg}, D.~H., {Yanny},
  B., {Yasuda}, N., Sep. 2000. {The Sloan Digital Sky Survey: Technical
  Summary}. \aj 120, 1579--1587.

\end{thebibliography}

\end{document}